%% file: rmptop.tex
\documentstyle[sprocl,epsf]{article}
\epsfverbosetrue

\def\Journal#1#2#3#4{{#1}{\bf #2}, #3 (#4)}

\def\NPB{{\em Nucl.~Phys.~}B}
\def\PLB{{\em Phys.~Lett.~}B}
\def\PRL{\em Phys.~Rev.~Lett.~}
\def\PRD{{\em Phys.~Rev.~}D}
\def\ZPC{{\em Z.~Phys.~}C}
\def\PTP{\em Prog.~Theor.~Phys.~}
\def\EPJ{{\em Eur.~Phys.~J.~}C}

\begin{document}

\title{Studying the Top Quark} 

\author{S.~WILLENBROCK}

\address{Department of Physics, University of Illinois at Urbana-Champaign, \\
1110 West Green Street, Urbana, IL 61801}


\maketitle\abstracts{The top quark, discovered at the Fermilab Tevatron
collider in 1995, is the heaviest known elementary particle.  Its large
mass suggests that it may play a special role in nature.  It behaves 
differently from the other known quarks due both to its large mass and its
short lifetime.  Thus far we have only crude measurements of the properties 
of the top quark, such as its mass, weak interactions, strong interactions, 
and decay modes.  These measurements will be made more precise when the 
Tevatron begins operation again in 2001.  I review the present status of these
measurements, and discuss their anticipated improvement.}

\section{Introduction}

There are six known quarks in nature, with the whimsical names up, down,
strange, charm, bottom, and top.  The quarks are arranged in three 
pairs or ``generations'', as shown in Fig.~1; each member of a pair may be 
transformed into its partner via the charged-current weak interaction.  
Together with the six known leptons
(the electron, muon, tau, and their associated neutrinos; see Fig.~1), 
the six quarks constitute
all of the matter\footnote{The quarks and leptons are spin 1/2 fermions;
it is customary to reserve the term ``matter'' for these particles.
The other known particles are spin 1 gauge bosons, which 
mediate forces between the quarks and leptons; the photon ($\gamma$) mediates
the electromagnetic interaction, the gluon ($g$) the strong interaction,
$W$ the charged-current weak interaction, and $Z$ the neutral-current
weak interaction.}
in the universe (with the possible exception of the mysterious 
``dark matter'').  It is therefore essential that we understand the properties 
of the quarks and leptons in detail.  

\begin{figure}[h]
\include{fermions}

\caption{The six known quarks are arranged in three generations.  
Each quark is transformed into its partner via the charged-current weak 
interaction.  The same is true of the six known leptons.}
\end{figure}

The most recently-discovered of the quarks and leptons is the top quark,
which was discovered in 1995 by the CDF~\cite{CDFTOP} and D0~\cite{D0TOP}
experiments at the Fermilab
Tevatron, a proton-antiproton collider of center-of-mass 
energy\footnote{The upper-case ``Mandelstam variable'' 
$S$ corresponds to the square of the total energy of the colliding 
proton and antiproton in the center-of-mass frame.}
$\sqrt S = 1.8$ TeV located in the suburbs of Chicago.  Due to its relatively
recent discovery, far less is known about the top quark than about the 
other quarks and leptons.  In this article I review what has been learned
about the top quark since its discovery (reviewed in this journal in 
Ref.~\cite{CF}),\footnote{For a non-technical exposition on the 
discovery of the top quark, see Ref.~\cite{LT}.} 
and look forward to future experimental probes of the top quark
at the Tevatron.\footnote{In this article
I restrict my attention to top-quark physics at the Tevatron.  The Tevatron 
has a monopoly on the top quark until 2005, when the CERN Large Hadron Collider
(a proton-proton collider of center-of-mass
energy $\sqrt S = 14$ TeV in Geneva, Switzerland) is scheduled to begin 
operation.  Proposed high-energy lepton colliders would also contribute to
top-quark physics.}

Thus far, the properties of the quarks and leptons are successfully described 
by the so-called ``standard model'' of the strong and electroweak interactions.
However, this theory does not account for the masses of these particles; it
merely accomodates them.  The top quark is by far the heaviest of the quarks
and leptons, and it is tempting to speculate that it is 
special.\footnote{Speculations about the special role of the top quark in
particle physics, and their experimental implications, are reviewed in 
Ref.~\cite{S}.} The goal
of future experiments is therefore to measure the properties of the top quark,
to compare them with the standard model, and to learn whether the top
quark is indeed special.  

What are the chances that a close inspection of the properties of the 
top quark will yield surprises?  One way to address this question is to 
consider the top-quark's weak-interaction partner, the $b$ quark.  
The $b$ quark
was discovered in 1977,\cite{BDISC} and in 1983 it yielded its first surprise:
its lifetime was found to be much longer than expected.\cite{MAC,MARKII}  
The top quark has already yielded its first surprise: the large value of 
its mass, approximately 174 GeV.  The next heaviest quark is the $b$ quark,
with a mass of only about 5 GeV. Fifteen years ago, there were few
who would have guessed that the top quark would be so heavy.  
A detailed scrutiny of
the top-quark's properties will reveal whether there are more surprises 
in top-quark physics.

Even if the top quark should prove to
be a normal quark, the experimental consequences of this very heavy quark
are interesting in their own right.  Many of the measurements described in
this article have no analogue for the lighter quarks.  This is not just
a consequence of the large mass of the top quark, but also of its very short
lifetime.  In contrast to the lighter quarks, which are permanently 
confined in bound states with other 
quarks and antiquarks,\footnote{These bound states, collectively called
hadrons, come is two types: baryons (three quarks) and mesons 
(quark and antiquark).  They are formed by the strong interaction.}
the top quark decays so quickly that it does not have time to form bound 
states.  There is also insufficient time to depolarize the spin of the
top quark, in contrast to the lighter quarks, whose spin is depolarized
by chromomagnetic interactions\footnote{This is the strong-interaction 
analogue of magnetism.} within the bound states.  Thus the top quark is
free of many of the complications associated with the strong interaction.
The top quark therefore
presents novel experimental challenges and opportunites,
which require innovative ideas and techniques.

\section{Overview}

The top quark was discovered during Run I of the Tevatron,\cite{CDFTOP,D0TOP} 
from 1992-1996, in which approximately 100~pb$^{-1}$ of integrated 
luminosity\footnote{The integrated luminosity corresponds
to the instantaneous luminosity integrated over time.} 
were collected.
The top quark is believed to have a very short lifetime,
about $0.5 \times 10^{-24}$ s, so it can only be detected indirectly via
its decay products, a $W$ boson and a $b$ quark ($t \to Wb$).  
A $b$ quark is sufficiently long-lived (1.5~ps) that it travels a measurable
distance before decaying (about 450 $\mu$m), leaving a secondary vertex
which can be detected with a silicon vertex detector 
(``$b$ tagging'').\footnote{A $b$ quark can also be tagged
via its semileptonic decay.}  The $W$ boson 
can decay either to a pair of leptons or a pair of quarks.  The top quark is
produced via the strong interaction together with its antiparticle, 
the top antiquark (denoted $\bar t$).

Run II of the Tevatron is scheduled to begin in 2001, with an initial
goal of 2~fb$^{-1}$ of integrated luminosity, and an ultimate goal
of up to 30~fb$^{-1}$.  The machine energy in Run II will be 
$\sqrt S = 2$ TeV, an increase over the $\sqrt S = 1.8$ TeV energy of Run I.  
Both the CDF and D0 detectors will be upgraded such that
they will have an increased acceptance 
for top-quark events.\cite{CDFRUNII,D0RUNII}

These improvements in the accelerator and detectors translate
into a large number of top quarks. 
For example, let's consider some of the cleanest
top-quark events, $t\bar t \to WWb\bar b$, where one $W$ boson is detected 
via its leptonic decay, the other $W$ boson decays to a pair of quarks,
and at least one of the $b$ quarks is tagged.  These events are 
fully reconstructable and have very little background.  In the Run I data, 
each experiment had about 25 such 
events.\cite{CDFMTOP,D0MTOP}\footnote{For example, CDF had 34 such events, 
of which about 8 are thought to be background.}
There are expected to be about 1000 events per experiment in the initial 
stage of Run II (2~fb$^{-1}$), due mostly to the factor of 20 increase in 
integrated luminosity, but also due to the 37\% increase in production 
cross section at 
$\sqrt S = 2$ TeV and the increased acceptance for top-quark events.  
The ultimate goal of 30~fb$^{-1}$ corresponds
to about 15,000 events per experiment.  The large number of events 
produced in Run II will allow a detailed scrutiny of the properties 
of the top quark.

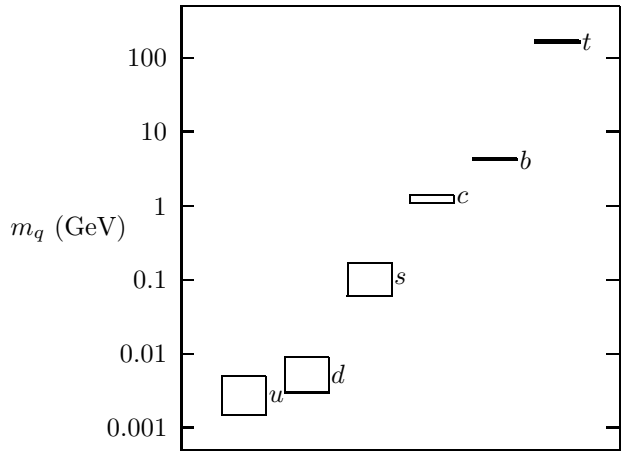
\begin{figure}[t]
\begin{center}
\input{massspectrum}
\end{center}
\caption{The quark mass spectrum.  The bands indicate the  
running $\overline{\rm MS}$ mass, evaluated at the quark mass 
(for $c,b,t$) or 
at 2 GeV (for $u,d,s$), and the associated uncertainty.}  
\label{fig:massspectrum}
\end{figure}

\section{Top mass}

The top-quark mass has been measured by the CDF~\cite{CDFMASS} and 
D0~\cite{D0MASS} collaborations to be
\begin{eqnarray}
m_t & = & 176.0 \pm 6.5\;{\rm GeV}\;({\rm CDF}) \\ 
    & = & 172.1 \pm 7.1\;{\rm GeV}\;({\rm D0})\;. 
\end{eqnarray}
This yields a world-average mass of~\cite{AVGMASS}\footnote{This is the 
top-quark pole mass, which corresponds approximately to its 
physical mass.\cite{SmW} The corresponding $\overline{\rm MS}$ mass, which
is an unphysical parameter useful for precision analyses, is 
$m_t^{\overline {\rm MS}}(m_t^{\overline {\rm MS}})
= 165.2 \pm 5.1$ GeV.\cite{GBGS}}
\begin{equation}
m_t = 174.3 \pm 5.1\;{\rm GeV}\;({\rm CDF} + {\rm D0})\;.
\end{equation}
To put this into context, I plot all the quark masses in Fig.~2, 
on a logarithmic scale. The width
of each band is proportional to the fractional uncertainty in the quark mass.  
We see that, at present, the top-quark mass is the best-known quark mass,
with the $b$-quark mass a close second 
($m_b^{\overline {\rm MS}}(m_b)
=4.25 \pm 0.15$ GeV).\cite{PDG}

\begin{figure}[t]
\begin{center}
\epsfxsize= 3.0in
\leavevmode
\epsfbox{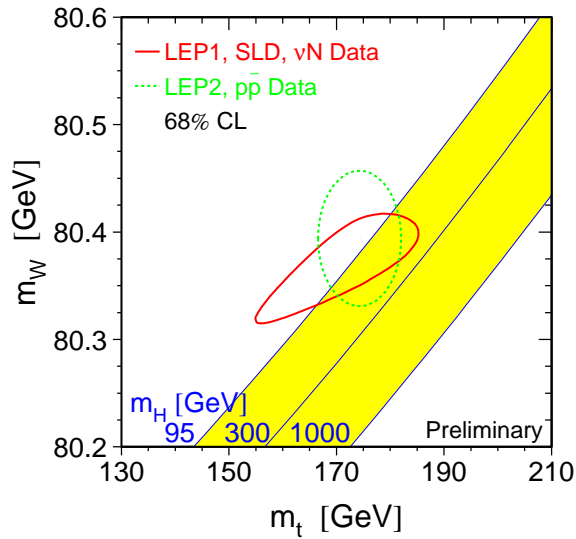}
\end{center}
\caption[fake]{$W$ mass {\it vs.}~top-quark mass, with lines of 
constant Higgs mass.
The solid ellipse is the $1\sigma$ ($68\%$ CL) contour from precision 
electroweak experiments.  The dashed ellipse is the $1\sigma$ ($68\%$ CL)
contour from direct measurements.  Only the shaded region is allowed in the 
standard electroweak model.  Figure from Ref.~\cite{LEPEWWG}.}
\end{figure}

An important question for the future is what precision we desire for the 
top-quark mass.  There are at least two avenues along which to address this
question.  One is in the context of precision electroweak data.
Fig.~3 summarizes the world's precision electroweak data on a plot of 
$M_W$ {\it vs.}~$m_t$.  
The solid ellipse is the $1\sigma$ contour.  If the
standard electroweak model is correct, the measured top-quark mass should
lie within this contour.  Since the contour spans about $\pm 8$ GeV 
along the $m_t$ axis, we conclude that the present uncertainty of $5$ GeV
in the top-quark mass is more than sufficient for the purpose of precision
electroweak physics at this time.  

There is one electroweak
measurement, $M_W$, whose precision could increase significantly.
An uncertainty of $20$ MeV is a realistic goal for Run II (30~fb$^{-1}$)
at the Tevatron.\cite{TEV2000}
Let us take this uncertainty and project it onto
a line of constant Higgs mass in Fig.~3.\footnote{The hypothetical Higgs boson
is discussed in the Outlook.}  This is appropriate, 
because once a Higgs boson is discovered, even a crude knowledge of
its mass will define a narrow line in Fig.~3, since precision electroweak
measurements are sensitive only to the logarithm of the Higgs mass.  
An uncertainty in $M_W$ of $20$ MeV projected onto a line of constant
Higgs mass corresponds to an uncertainty of $3$ GeV in the top-quark
mass.  Thus we desire a measurement of $m_t$ to $3$ GeV 
in order to make maximal use of the precision measurement of $M_W$.

\begin{figure}[t]
\begin{center}
\setlength{\unitlength}{1.0in}
\begin{picture}(3.0,1.5)(0.0,0.0)
\epsfxsize= 1.5in
\leavevmode
\put(0.0,0.0){\epsfbox{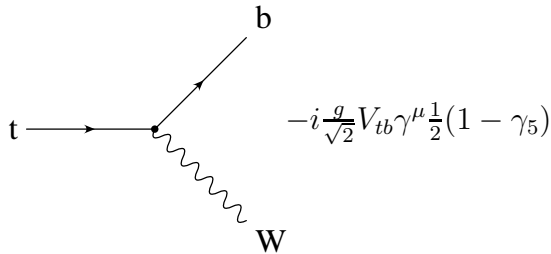}}
\put(1.5,0.88){{\large $-i \frac{g}{\sqrt{2}} V_{tb} \gamma^{\mu} 
\frac{1}{2}(1-\gamma_{5})$}}
\end{picture}
\caption{Top-quark charged-current weak interaction.}
\end{center}
\end{figure}

Another avenue along which to address the desired accuracy of the top-quark 
mass is to
recall that the top-quark mass is a fundamental parameter of the standard
model.  Actually, the fundamental parameter is the coupling of the 
top quark to the Higgs field (``Yukawa coupling''), given by
\begin{equation}
y_t = \sqrt 2 \frac{m_t}{v} \approx 1\
\label{yukawa}
\end{equation}
where $v \approx 246$ GeV is the vacuum-expectation value of the Higgs field.
The fact that this coupling is of order unity suggests that it may be a
truly fundamental parameter.  We hope someday to have a theory that relates
the top-quark Yukawa coupling to that of its weak-interaction partner, 
the $b$ quark.\footnote{A particularly compelling model which relates
the $b$ and $t$ masses is SO(10) grand 
unification.\cite{G,FM} This model may be able to account for the 
masses of all the third-generation fermions, including the tau neutrino,
whose mass is given by the ``see-saw'' mechanism \cite{GRS} as
$m_{\nu_\tau} \approx m_t^2/M_{GUT} \approx 10^{-2}$ eV.\cite{W}} 
The $b$-quark mass is currently known with an accuracy of $3.5\%$.
Since the uncertainty is entirely theoretical, it is likely that it will
be reduced in the future.  If we assume that future work cuts the uncertainty 
in half, the corresponding uncertainty in the top-quark mass would be
$3$ GeV.

We conclude that both precision electroweak experiments and $m_t$ as a 
fundamental parameter lead us to the desire to measure the top-quark
mass with an accuracy of $3$ GeV.  This is well matched with 
future expectations.  An uncertainty of $3$ GeV per experiment 
is anticipated in the initial stage of 
Run II (2~fb$^{-1}$),\cite{CDFRUNII,D0RUNII} and additional running 
could reduce this uncertainty to $2$ GeV.\cite{TEV2000}

\section{Top weak interaction}

The standard model dictates that the top quark has the same 
vector-minus-axial-vector ($V-A$) charged-current weak interaction 
as all the other fermions, as shown in Fig.~4.
It is easy to see that this implies that the $W$ boson in top decay 
cannot be right handed, {\it i.e.}, have positive 
helicity.\footnote{Helicity is
the component of spin along the direction of motion of a particle.}   
The argument is sketched in
Fig.~5.  In the idealized limit of a massless $b$ quark, 
the $V-A$ current dictates
that the $b$ quark in top decay is always left-handed.\footnote{Being far
from massless, the decaying top quark can be left- or right-handed.}
If the $W$ boson were right-handed, then the component of total angular
momentum along the decay axis would be $+3/2$ (there is no component
of orbital angular momentum along this axis).  But the initial top quark
has spin angular momentum $\pm 1/2$ along this axis, so this decay is
forbidden by conservation of angular momentum.  
CDF has measured
\begin{equation}
BR(t\to W_+b) = 0.11 \pm 0.15
\end{equation}
which is consistent with zero.\cite{HELICITY}

\begin{figure}[t]
\begin{center}
\epsfxsize= 2.1in
\leavevmode
\epsfbox{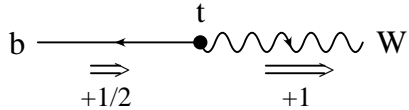}
\end{center}
\caption{Illustration that the top quark cannot decay to a right-handed
(positive-helicity) $W$ boson.}
\end{figure}

The top quark may decay to a left-handed (negative helicity) or a longitudinal
(zero helicity) $W$ boson.
Its coupling to a longitudinal $W$ boson is similar to its Yukawa coupling,
Eq.~(\ref{yukawa}), which is enhanced with respect to the weak coupling.
Therefore the top quark prefers to decay to 
a longitudinal $W$ boson, with a branching ratio
\begin{equation}
BR(t\to W_0 b) = \frac{m_t^2}{m_t^2+2M_W^2} \approx 70\%\;.
\end{equation}
CDF has made a first measurement of this branching ratio,\cite{HELICITY} 
\begin{equation}
BR(t\to W_0 b) = 0.91 \pm 0.37 \pm 0.13\;,
\end{equation}
which is consistent with expectations.
The anticipated fractional accuracy of this measurement in the 
initial stage of Run II 
(2 fb$^{-1}$) is $5.5\%$,\cite{CDFRUNII,D0RUNII} with an ultimate accuracy 
(30 fb$^{-1}$) of less than $2\%$.\cite{TEV2000}  

\begin{figure}[t]
\begin{center}
\epsfxsize= 3.8in
\leavevmode
\epsfbox{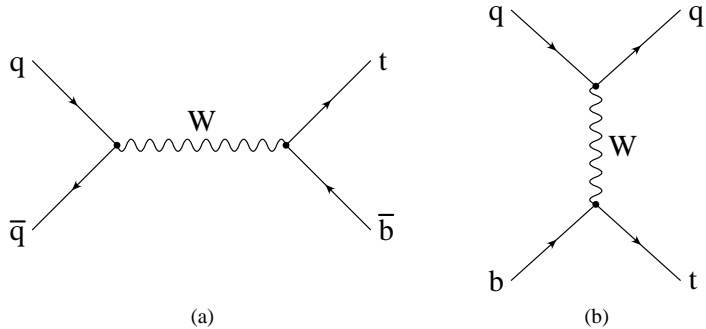}
\end{center}
\caption{Single-top-quark production via the weak interaction: 
(a) $s$-channel process; (b) $t$-channel process.}
\end{figure}

Quarks are transformed into their partner via the charged-current weak
interaction, but they are not completely loyal; there are also 
occasional transitions
between different generations.  This is described by the $3 \times 3$
Cabibbo-Kobayashi-Maskawa (CKM) matrix.  The matrix elements 
$V_{tb}$, $V_{ts}$, and $V_{td}$ characterize the strength of the 
transition of a top quark into a bottom, strange, and down quark, respectively
($|V_{ij}|\le 1$).

CDF has measured \cite{TOLLEFSON}
\begin{equation}
\frac{BR(t\to Wb)}{BR(t\to Wq)} 
= \frac{|V_{tb}|^2}{|V_{td}|^2+|V_{ts}|^2+|V_{tb}|^2} = 0.99 \pm 0.29
\label{ratio}
\end{equation}
and it is interesting to ask what this tells us about
$V_{tb}$.  If we assume that there are just three generations of quarks,
then unitarity of the CKM matrix implies that the denominator of 
Eq.~(\ref{ratio}) is unity, and we can immediately extract
\begin{equation}
|V_{tb}| = 0.99 \pm 0.15 \;(> 0.76\;{\rm 95\%\;CL}) \;(3\;{\rm generations}).
\end{equation}
However, to put this into perspective, recall that three-generation 
unitarity also implies that 
$|V_{ub}|^2+|V_{cb}|^2+|V_{tb}|^2=1$, and since $|V_{ub}|$ and $|V_{cb}|$
have been measured to be small, one finds \cite{PDG}
\begin{equation}
|V_{tb}| = 0.9991 - 0.9994 \;(3\;{\rm generations})
\end{equation}
which is far more accurate than the present CDF result (as well as
the anticipated accuracy from Run II).

If we assume more than three generations, then unitarity implies almost
nothing about $|V_{tb}|$:~\cite{PDG}
\begin{equation}
|V_{tb}| = 0.06 - 0.9994\;(>3\;{\rm generations})\;.
\end{equation}
At the same time, we also lose the constraint that the denominator of
the middle expression in Eq.~(\ref{ratio}) is unity.  
All we can conclude from Eq.~(\ref{ratio})
is that $|V_{tb}| >> |V_{ts}|,|V_{td}|$; we learn nothing about its
absolute magnitude.  

Fortunately, there is a direct way to measure $|V_{tb}|$ at the Tevatron, 
which makes no assumptions about the number of generations.  One uses
the weak interaction to produce the top quark; the two relevant processes
are shown in Fig.~6.  The cross sections for these two ``single top''
processes are proportional to $|V_{tb}|^2$.  The first process
involves an $s$-channel $W$ boson,\cite{CP,SW1,SmWNLO,HBB,SSW,BBD} while
the second process involves
a $t$-channel $W$ boson\footnote{It is conventional to label the Feynman
diagrams by the lower-case ``Mandelstam variables''
$s$ and $t$, which correspond to 
the square of the four-momentum of the $W$ boson in the diagrams.}
(and is often called $W$-gluon fusion, because
the initial $b$ quark actually comes from a gluon splitting 
to $b\bar b$).\cite{DW,Y,EP,CY,HBB,SSWNLO,SSW,BBD}
The $t$-channel process has the advantage of greater statistics than the 
$s$-channel process,
but the disadvantage of greater theoretical uncertainty.  
Thus far there is only a bound on single-top-quark production via
the $t$-channel process from CDF,\cite{TOLLEFSON} 
\begin{equation}
\sigma(qb\to qt) < 15.4\;{\rm pb}\;({\rm 95\%\;CL})
\end{equation}
which is an order of magnitude away from the theoretical expectation 
of $1.70 \pm 0.24$~pb.\cite{SSWNLO,SSW}  There is a similar bound on 
the $s$-channel process from CDF,\cite{SAVARD}
\begin{equation}
\sigma(q\bar q \to t\bar b) < 15.8\;{\rm pb}\;({\rm 95\%\;CL})
\end{equation}
which is even further from the theoretical expectation of 
$0.73 \pm 0.10$~pb.\cite{SmWNLO}

Both single-top processes should 
be observed in the initial stage of Run II (2~fb$^{-1}$); 
the $t$-channel process will yield a measurement of $V_{tb}$ with an accuracy
of about $13\%$.\cite{CDFRUNII,D0RUNII}  The ultimate accuracy (30 fb$^{-1}$)
is anticipated to be about $5\%$, perhaps using the $s$-channel 
process owing to its small theoretical uncertainty.\cite{TEV2000}

\begin{figure}[t]
\begin{center}
\epsfxsize= 4.0in
\leavevmode
\epsfbox{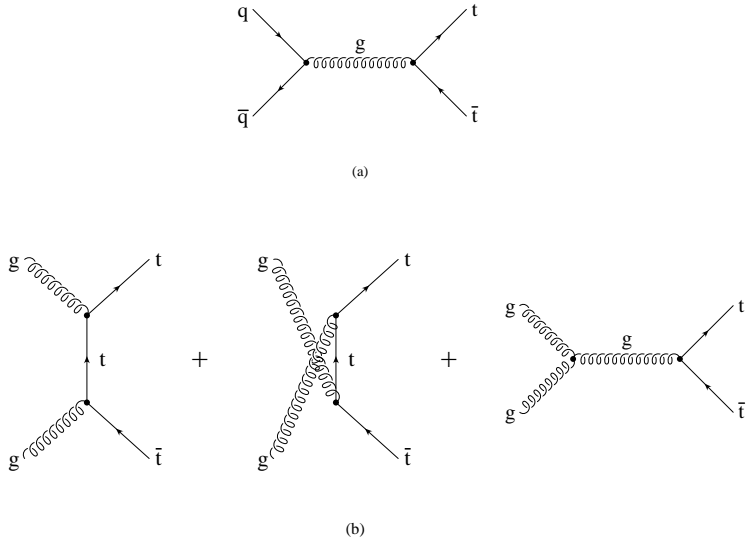}
\end{center}
\caption{Top-quark pair production via the strong interaction:
(a) quark-antiquark annihilation; (b) gluon fusion. There are three Feynman
diagrams which contribute to the latter process.}
\end{figure}

Single-top-quark production can also be used to test the $V-A$ structure
of the top-quark charged-current weak interaction.  This structure implies that
the top-quark spin is nearly 100\% polarized along the direction (in the 
top-quark rest frame) of the $d$ or $\bar d$ quark in the event, 
in both $W$-gluon fusion and the $s$-channel process.\cite{MP}
This effect will be observable in Run II.\cite{SSW}

\section{Top strong interaction}

The strong interaction of the top quark is best tested in its production.
There are two subprocesses by which $t\bar t$ pairs are produced via the 
strong interaction at a 
hadron collider, shown in Fig.~7.  At the Tevatron, the quark-antiquark
annihilation process is dominant, accounting for $90\%$ of the cross 
section at $\sqrt S = 1.8$ TeV.  When the machine energy is increased
to $\sqrt S = 2$ TeV in Run II, this fraction decreases to $85\%$.  
The cross section increases considerably, by about $37\%$, when the
machine energy is increased from 1.8 to 2 TeV.

\begin{figure}[t]
\begin{center}
\epsfysize=6.5 cm
\leavevmode
\epsfbox[35 150 530 655]{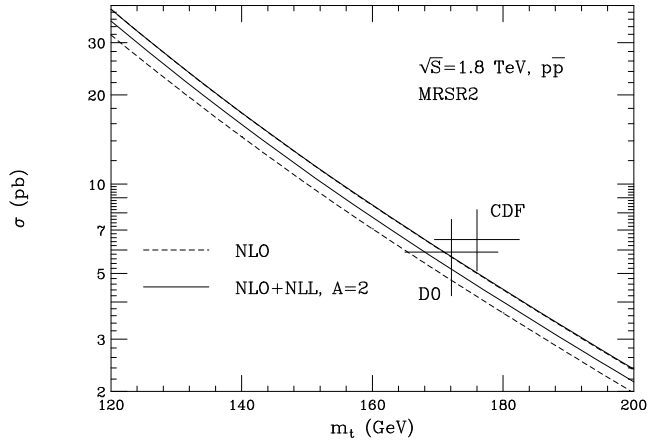} 
\end{center}
\caption[fake]{Cross section for $t\bar t$ production at the Tevatron 
{\it vs.}~the top-quark mass.  Dashed band is from next-to-leading-order (NLO)
in the strong interaction (note that the upper solid and dashed lines are 
nearly coincident);
solid band includes soft-gluon resummation at next-to-leading-logarithm (NLL).
The calculation employs the MRSR2 parton distribution functions to describe 
the quark and gluon content of the proton.\cite{MRSR2}
Figure adapted from Ref.~\cite{BCMN}.}
\label{fig:topcross}
\end{figure}

We show in Fig.~8 the $t\bar t$ cross section {\it vs.}~the top-quark mass.
The dashed band is from a calculation at next-to-leading-order (NLO)
in the strong interaction.\cite{NDE,BKNS} 
The uncertainty in this calculation is about $10\%$.
The solid band includes the effect of soft gluon resummation at 
next-to-leading logarithm (NLL); this increases the cross section by only a few
percent, but reduces the uncertainty by almost 
a factor of two.\cite{LSN,BC,CMNT,BCMN}\footnote{These bands reflect the 
uncertainty in the cross section 
due to the variation of the renormalization and factorization scales.
They do not include the uncertainty
from $\alpha_s(M_Z)$ or the parton distribution functions.  However, these
additional uncertainties are relatively modest.\cite{CMNT}}    
The measurements by CDF~\cite{CDFCROSS} and D0,\cite{D0CROSS}
\begin{eqnarray}
\sigma & = & 6.5^{+1.7}_{-1.4}\; {\rm pb}\;({\rm CDF}) \\
\sigma & = & 5.9 \pm 1.7\; {\rm pb}\;({\rm D0})
\end{eqnarray}
are also shown in the figure, and are seen to agree with theory within 
one standard deviation.  The anticipated accuracy of the measurement of
the cross section in the initial stage of 
Run II (2 fb$^{-1}$) is $9\%$,\cite{CDFRUNII,D0RUNII}
with an ultimate accuracy (30 fb$^{-1}$) of $5\%$.\cite{TEV2000}

\begin{figure}[t]
\begin{center}
\epsfxsize= 4.0in
\leavevmode
\epsfbox{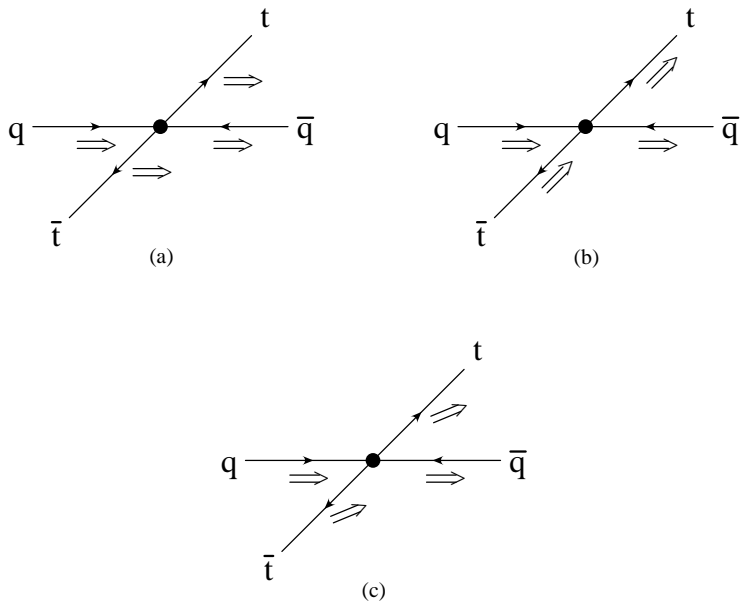}
\end{center}
\caption{Top-quark and light-quark spins in $q\bar q \to t\bar t$:
(a) near threshold; (b) far above threshold; (c) intermediate energies.}
\end{figure}

An interesting aspect of the strong production of $t\bar t$ pairs is that
the spins of the $t$ and $\bar t$ are nearly 
$100\%$ correlated.\cite{K,BOP,H,AS,MP1,SW2,B,CLS,MP2}
The correct basis in which to measure the spins requires some consideration,
however.  At threshold ($\sqrt s \approx 2m_t$),\footnote{The lower-case 
``Mandelstam variable'' $s$ corresponds to the square of the total energy 
of the colliding quarks in the center-of-mass frame.} 
the cross section is
entirely $s$ wave, so the spins of the colliding quarks are transferred
to the $t$ and $\bar t$.  Since the quark-antiquark
annihilation takes place via a gauge interaction, the quark and antiquark
must have opposite helicities, so the spins of the $t$ and $\bar t$ 
are aligned along the beamline as shown in Fig.~9(a).  At the other
extreme, far above threshold ($\sqrt s >> 2m_t$), the $t$ and $\bar t$ 
behave like massless quarks, and therefore must have opposite helicities,
as shown in Fig.~9(b).
The question is whether there is a basis which interpolates between the 
beamline basis near threshold and the helicity basis far above threshold,
and the answer is affirmative - it has been dubbed the ``off-diagonal'' 
basis.\cite{PS,MP2}  The $t$ and $\bar t$ spins are $100\%$ correlated in this
basis, as shown in Fig.~9(c). 
Since the quark-antiquark annihilation process accounts for most
of the cross section at the Tevatron, the spin correlation is nearly $100\%$.
A first attempt to observe this effect has been made by D0, based on
six dilepton events.\cite{D0SPIN} 
This effect should be observable in Run II.\cite{SNYDER}

Another interesting aspect of the strong production of $t\bar t$ pairs is 
an asymmetry in the distribution of the $t$ and $\bar t$ quarks.\cite{KR}  
This effect arises at next-to-leading order, and leads to a 
forward-backward asymmetry 
of about 5\% in $t\bar t$ production at the Tevatron.

\section{Rare decays}

\begin{figure}[t]
\begin{center}
\epsfxsize= 3.40in
\leavevmode
\epsfbox{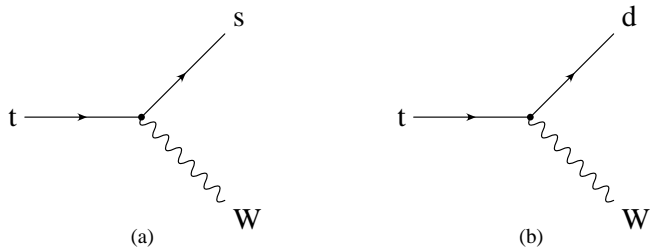}
\end{center}
\caption{Rare top decays: (a) $t\to Ws$; (b) $t\to Wd$.}
\end{figure}

Rare top decays in the standard model tend to be very rare, outside the
range of the Tevatron.  Thus far CDF has placed limits on the rare decays
\cite{TZC}
\begin{eqnarray}
BR(t\to Zq) & < & 33\%\;(95\%\;{\rm CL})\\
BR(t\to \gamma q) & < & 3.2\%\;(95\%\;{\rm CL})
\end{eqnarray}
which have tiny branching ratios in the standard model.\cite{EHS}

The least rare of the rare decays within the standard model are the CKM 
suppressed decays $t\to Ws$ and $t\to Wd$, shown in Fig.~10.  These decays
are interesting because they allow a direct measurement of the CKM matrix
elements $V_{ts}$ and $V_{td}$.  Assuming three generations, the branching 
ratios are predicted to be
\begin{eqnarray}
BR(t\to Ws) & \approx & 0.1\% \\
BR(t\to Wd) & \approx & 0.01\%
\end{eqnarray}
which are small, but not tiny.  Since there will be about 10,000 raw $t\bar t$
pairs produced in the initial stage of Run II (2~fb$^{-1}$), and about 
150,000 ultimately (30~fb$^{-1}$), events of these
types will be present in the data.  However, there is no 
generally-accepted strategy for identifying these events.

\section{Outlook}

Top-quark physics is in its infancy.  Since its discovery in 1995, we have 
only had a crude look at the top-quark's properties.  
Run II of the Fermilab Tevatron,
scheduled to begin in 2001, will allow a careful study of the top quark;
its strong and weak interactions, as well as its mass, will be
accurately measured.  The goal of these studies is to determine if the top
quark, which is so much heavier than the other quarks and leptons, is
special.  Even if the top quark should prove to be normal, the study of
this very massive quark will be intriguing, since many of these studies
have no analogue for the lighter quarks.

The CERN Large Hadron Collider, scheduled to begin
operation in 2005, will allow an even closer look at the 
top quark.\cite{ATLAS,YELLOW}
Proposed lepton colliders would provide a complementary view of the top
quark, especially if they are designed to operate near the $t\bar t$
threshold.\cite{MuP}

It is not known if the study of the top quark will bring us closer to 
an understanding of the mechanism which endows the top quark, as well as 
the other quarks and leptons, with mass.  In the standard Higgs model,
it is the coupling of the quarks and leptons to the Higgs field which 
is responsible for the generation of mass.\footnote{This coupling
also yields the Cabbibo-Kobayashi-Maskawa matrix.} The discovery of the Higgs
boson would be compelling evidence that this model is correct in its
essence.  The search for the Higgs boson, or whatever else Nature has 
provided, is a central focus of particle physics.  If we are 
fortunate, the Higgs boson could be discovered in Run II of the 
Tevatron.\cite{SMW,TEV2000,HZ,HTZ,BHT,CMW}  The Higgs boson cannot elude the 
Large Hadron Collider,\cite{ATLAS} so we are certain to glean important 
information about the generation of mass in the coming decade.  

\section*{Acknowledgments}

I am grateful for conversations with and assistance from T.~Liss, M.~Mangano,
K.~Paul, R.~Roser, S.~Snyder, and T.~Stelzer.
This work was supported in part by Department of Energy grant 
DE-FG02-91ER40677.

\end{document}

%% file: fermions.tex
\begin{center}
\begin{tabular}{lccc}
Quarks:  & $\left( \begin{array}{c}
                      u \\ d
                   \end{array}
            \right)$
         & $\left( \begin{array}{c} 
                      c \\ s
                   \end{array}
            \right)$
         & $\left( \begin{array}{c}
                      t \\ b 
                   \end{array}
            \right)$ \\
\\
Leptons: & $\left( \begin{array}{c}
                      \nu_{e} \\ e 
                   \end{array}
            \right)$
         & $\left( \begin{array}{c}
                      \nu_{\mu} \\ \mu
                   \end{array}
            \right)$
         & $\left( \begin{array}{c}
                      \nu_{\tau} \\ \tau
                   \end{array}
            \right)$
\end{tabular}
\end{center}

%% file: massspectrum.tex
\setlength{\unitlength}{0.240900pt}
\ifx\plotpoint\undefined\newsavebox{\plotpoint}\fi
\sbox{\plotpoint}{\rule[-0.200pt]{0.400pt}{0.400pt}}%
\begin{picture}(974,809)(0,0)
\font\gnuplot=cmr10 at 10pt
\gnuplot
\sbox{\plotpoint}{\rule[-0.200pt]{0.400pt}{0.400pt}}%
\put(221.0,103.0){\rule[-0.200pt]{4.818pt}{0.400pt}}
\put(199,103){\makebox(0,0)[r]{0.001}}
\put(890.0,103.0){\rule[-0.200pt]{4.818pt}{0.400pt}}
\put(221.0,219.0){\rule[-0.200pt]{4.818pt}{0.400pt}}
\put(199,219){\makebox(0,0)[r]{0.01}}
\put(890.0,219.0){\rule[-0.200pt]{4.818pt}{0.400pt}}
\put(221.0,335.0){\rule[-0.200pt]{4.818pt}{0.400pt}}
\put(199,335){\makebox(0,0)[r]{0.1}}
\put(890.0,335.0){\rule[-0.200pt]{4.818pt}{0.400pt}}
\put(221.0,451.0){\rule[-0.200pt]{4.818pt}{0.400pt}}
\put(199,451){\makebox(0,0)[r]{1}}
\put(890.0,451.0){\rule[-0.200pt]{4.818pt}{0.400pt}}
\put(221.0,568.0){\rule[-0.200pt]{4.818pt}{0.400pt}}
\put(199,568){\makebox(0,0)[r]{10}}
\put(890.0,568.0){\rule[-0.200pt]{4.818pt}{0.400pt}}
\put(221.0,684.0){\rule[-0.200pt]{4.818pt}{0.400pt}}
\put(199,684){\makebox(0,0)[r]{100}}
\put(890.0,684.0){\rule[-0.200pt]{4.818pt}{0.400pt}}
\put(221.0,68.0){\rule[-0.200pt]{165.980pt}{0.400pt}}
\put(910.0,68.0){\rule[-0.200pt]{0.400pt}{167.907pt}}
\put(221.0,765.0){\rule[-0.200pt]{165.980pt}{0.400pt}}
\put(45,416){\makebox(0,0){$m_{q}$ (GeV)}}
\put(359,153){\makebox(0,0)[l]{$u$}}
\put(457,187){\makebox(0,0)[l]{$d$}}
\put(556,338){\makebox(0,0)[l]{$s$}}
\put(654,465){\makebox(0,0)[l]{$c$}}
\put(753,524){\makebox(0,0)[l]{$b$}}
\put(851,709){\makebox(0,0)[l]{$t$}}
\put(221.0,68.0){\rule[-0.200pt]{0.400pt}{167.907pt}}
\put(285,123){\usebox{\plotpoint}}
\put(285.0,123.0){\rule[-0.200pt]{16.622pt}{0.400pt}}
\put(354.0,123.0){\rule[-0.200pt]{0.400pt}{14.695pt}}
\put(285.0,184.0){\rule[-0.200pt]{16.622pt}{0.400pt}}
\put(285.0,123.0){\rule[-0.200pt]{0.400pt}{14.695pt}}
\put(383,158){\usebox{\plotpoint}}
\put(383.0,158.0){\rule[-0.200pt]{16.622pt}{0.400pt}}
\put(452.0,158.0){\rule[-0.200pt]{0.400pt}{13.490pt}}
\put(383.0,214.0){\rule[-0.200pt]{16.622pt}{0.400pt}}
\put(383.0,158.0){\rule[-0.200pt]{0.400pt}{13.490pt}}
\put(481,310){\usebox{\plotpoint}}
\put(481.0,310.0){\rule[-0.200pt]{16.863pt}{0.400pt}}
\put(551.0,310.0){\rule[-0.200pt]{0.400pt}{12.527pt}}
\put(482.0,362.0){\rule[-0.200pt]{16.622pt}{0.400pt}}
\put(482.0,310.0){\rule[-0.200pt]{0.400pt}{12.527pt}}
\put(580,456){\usebox{\plotpoint}}
\put(580.0,456.0){\rule[-0.200pt]{16.622pt}{0.400pt}}
\put(649.0,456.0){\rule[-0.200pt]{0.400pt}{2.891pt}}
\put(580.0,468.0){\rule[-0.200pt]{16.622pt}{0.400pt}}
\put(580.0,456.0){\rule[-0.200pt]{0.400pt}{2.891pt}}
\put(679,523){\usebox{\plotpoint}}
\put(679.0,523.0){\rule[-0.200pt]{16.622pt}{0.400pt}}
\put(748.0,523.0){\rule[-0.200pt]{0.400pt}{0.723pt}}
\put(679.0,526.0){\rule[-0.200pt]{16.622pt}{0.400pt}}
\put(679.0,523.0){\rule[-0.200pt]{0.400pt}{0.723pt}}
\put(777,708){\usebox{\plotpoint}}
\put(777.0,708.0){\rule[-0.200pt]{16.622pt}{0.400pt}}
\put(846.0,708.0){\rule[-0.200pt]{0.400pt}{0.723pt}}
\put(777.0,711.0){\rule[-0.200pt]{16.622pt}{0.400pt}}
\put(777.0,708.0){\rule[-0.200pt]{0.400pt}{0.723pt}}
\end{picture}

%% file: rmptop.bbl
\section*{References}

\begin{thebibliography}{99}

\bibitem{CDFTOP} CDF Collaboration, F.~Abe {\it et al.}, 
\Journal{\PRL}{74}{2626}{1995}.

\bibitem{D0TOP} D0 Collaboration, S.~Abachi {\it et al.},
\Journal{\PRL}{74}{2632}{1995}.

\bibitem{CF} C.~Campagnari and M.~Franklin, {\em Rev.~Mod.~Phys.~}{\bf 69}, 
137 (1997). 

\bibitem{LT} T.~Liss and P.~Tipton, {\em Sci.~Am.~}{\bf 277}, No.~3, 54 (1997).

\bibitem{S} E.~Simmons, hep-ph/9908511.

\bibitem{BDISC} S.~Herb {\it et al.}, \Journal{\PRL}{39}{252}{1977}.

\bibitem{MAC} MAC Collaboration, E.~Fernandez {\it et al.}, 
\Journal{\PRL}{51}{1022}{1983}. 

\bibitem{MARKII} Mark II Collaboration, N.~Lockyer {\it et al.}, 
\Journal{\PRL}{51}{1316}{1983}.

\bibitem{CDFRUNII} {\sl CDF II Technical Design Report}, FERMILAB-Pub-96/390-E
(1996). 

\bibitem{D0RUNII} {\sl The D0 Upgrade: The Detector and Its Physics}, 
FERMILAB-Pub-96/357-E (1996).

\bibitem{CDFMTOP} CDF Collaboration, F.~Abe {\it et al.}, 
\Journal{\PRL}{80}{2767}{1998}.

\bibitem{D0MTOP} D0 Collaboration, S.~Abachi {\it et al.},
\Journal{\PRL}{79}{1197}{1997}.

\bibitem{CDFMASS} CDF Collaboration, F.~Abe {\it et al.}, 
\Journal{\PRL}{82}{271}{1999}.

\bibitem{D0MASS} D0 Collaboration, B.~Abbott {\it et al.},
\Journal{\PRL}{80}{2063}{1998};
\Journal{\PRD}{58}{052001}{1998}; {\bf 60}, 052001 (1999).

\bibitem{AVGMASS} L.~Demortier, R.~Hall, R.~Hughes, B.~Klima, R.~Roser,   
and M.~Strovink, FERMILAB-TM-2084 (1999).

\bibitem{SmW} M.~Smith and S.~Willenbrock, \Journal{\PRL}{79}{3825}{1997}.

\bibitem{GBGS} N.~Gray, D.~Broadhurst, W.~Grafe, and K.~Schilcher,
\Journal{\ZPC}{48}{673}{1990}.

\bibitem{PDG} {\sl Review of Particle Physics}, Particle Data Group,
\Journal{\EPJ}{3}{1}{1998}.

\bibitem{LEPEWWG} LEP Electroweak Working Group, http://www.cern.ch/LEPEWWG/.

\bibitem{TEV2000} {\sl Future Electroweak Physics at the Fermilab Tevatron:
Report of the tev\_2000 Study Group}, eds.~D.~Amidei and R.~Brock, 
FERMILAB-Pub-96/082 (1996).

\bibitem{G} H.~Georgi, in {\sl Particles and Fields} 1974, ed. C. Carlson 
(AIP, New York, 1975), p. 575. 

\bibitem{FM} H.~Fritzsch and P. Minkowski, 
{\em Ann.~Phys.~}{\bf 93}, 193 (1975).

\bibitem{GRS} M.~Gell-Mann, P.~Ramond, and R.~Slansky, in {\sl Supergravity},
eds. P.~van Nieuwenhuizen and D.~Freedman (North Holland, Amsterdam, 1979),
p.~315.

\bibitem{W} F.~Wilczek, {\em Nucl.~Phys.~Proc.~Suppl.}~{\bf 77}, 511 (1999) 
[hep-ph/9809509].

\bibitem{HELICITY} CDF Collaboration, T.~Affolder {\it et al.}, 
\Journal{\PRL}{84}{216}{2000}.

\bibitem{TOLLEFSON} K.~Tollefson (CDF Collaboration), in {\sl Proceedings of
the 29th International Conference on High Energy Physics}, Vancouver,
Canada, July 23--29, 1998, eds.~A.~Astbury, D.~Axen, and J.~Robinson 
(World Scientific, Singapore, 1999), Vol.~II, p.~1112.

\bibitem{CP} S.~Cortese and R.~Petronzio, \Journal{\PLB}{253}{494}{1991}.

\bibitem{SW1} T.~Stelzer and S.~Willenbrock, \Journal{\PLB}{357}{125}{1996}.

\bibitem{SmWNLO} M.~Smith and S.~Willenbrock, \Journal{\PRD}{54}{6696}{1996}.

\bibitem{HBB} A.~Heinson, A.~Belyaev, and E.~Boos, 
\Journal{\PRD}{56}{3114}{1997}

\bibitem{SSW} T.~Stelzer, Z.~Sullivan, and S.~Willenbrock,
\Journal{\PRD}{58}{094021}{1998}.

\bibitem{BBD} A.~Belyaev, E.~Boos, and L.~Dudko, \Journal{\PRD}{59}{075001}
{1999}.

\bibitem{DW} S.~Willenbrock and D.~Dicus, \Journal{\PRD}{34}{155}{1986}.

\bibitem{Y} C.-P.~Yuan, \Journal{\PRD}{41}{42}{1990}.

\bibitem{EP} R.~K.~Ellis and S.~Parke, \Journal{\PRD}{46}{3785}{1992}.

\bibitem{CY} D.~Carlson and C.-P.~Yuan, \Journal{\PLB}{306}{386}{1993}.

\bibitem{SSWNLO} T.~Stelzer, Z.~Sullivan, and S.~Willenbrock,
\Journal{\PRD}{56}{5919}{1997}.

\bibitem{SAVARD} P.~Savard (CDF Collaboration), FERMILAB-CONF-99-174-E, 
to appear in the {\sl Proceedings of the 34th Rencontres de Moriond on 
QCD and Hadronic Interactions}, Les Arcs, France, March 20--27, 1999.

\bibitem{MP} G.~Mahlon and S.~Parke, \Journal{\PRD}{55}{7249}{1997}.

\bibitem{NDE} P.~Nason, S.~Dawson, and R.~K.~Ellis, \Journal{\NPB}{303}{607}
{1988}. 

\bibitem{BKNS} W.~Beenakker, H.~Kuijf, W.~van Neerven, and J.~Smith, 
\Journal{\PRD}{40}{54}{1989}.

\bibitem{LSN} E.~Laenen, J.~Smith, and W.~van Neerven, 
\Journal{\NPB}{369}{543}{1992}.

\bibitem{BC} E.~Berger and H.~Contopanagos, \Journal{\PLB}{361}{115}{1995};
\Journal{\PRD}{54}{3085}{1996};\Journal{\PRD}{57}{253}{1998}.

\bibitem{CMNT} S.~Catani, M.~Mangano, P.~Nason, and L.~Trentadue, 
\Journal{\PLB}{378}{329}{1996}; \Journal{\NPB}{478}{273}{1996}.

\bibitem{BCMN} R.~Bonciani, S.~Catani, M.~Mangano, and P.~Nason,
\Journal{\NPB}{529}{424}{1998}

\bibitem{MRSR2} A.~Martin, R.~Roberts, and W.~J.~Stirling, 
\Journal{\PLB}{387}{419}{1996}.

\bibitem{CDFCROSS} CDF Collaboration, F.~Abe {\it et al.}, 
\Journal{\PRL}{80}{2773}{1998}; F.~Ptohos, presented at the International
Europhysics Conference on High Energy Physics, Tampere, Finland, 
July 15--21, 1999.

\bibitem{D0CROSS} D0 Collaboration, S.~Abachi {\it et al.},
\Journal{\PRL}{79}{1203}{1997}.

\bibitem{K} J.~K\"uhn, \Journal{\NPB}{237}{77}{1984}.

\bibitem{BOP} V.~Barger, J.~Ohnemus, and R.~Phillips, 
{\em Int.~J.~Mod.~Phys.~}{\bf A4}, 617 (1989).

\bibitem{H} Y.~Hara, \Journal{\PTP}{86}{779}{1991}.

\bibitem{AS} T.~Arens and L.~Sehgal, \Journal{\PLB}{302}{501}{1993}.

\bibitem{MP1} G.~Mahlon and S.~Parke, \Journal{\PRD}{53}{4886}{1996}.

\bibitem{SW2} T.~Stelzer and S.~Willenbrock, \Journal{\PLB}{374}{169}{1996}.

\bibitem{B} A.~Brandenburg, \Journal{\PLB}{388}{626}{1996}.

\bibitem{CLS} D.~Chang, S.-C.~Lee, and A.~Sumarokov, 
\Journal{\PRL}{77}{1218}{1996}.

\bibitem{MP2} G.~Mahlon and S.~Parke, \Journal{\PLB}{411}{173}{1997}.

\bibitem{PS} S.~Parke and Y.~Shadmi, \Journal{\PLB}{387}{199}{1996}.

\bibitem{D0SPIN} D0 Collaboration, B.~Abbott {\it et al.}, hep-ex/0002058.

\bibitem{SNYDER} S.~Snyder (D0 Collaboration), FERMILAB-CONF-99-294-E,
presented at the International
Europhysics Conference on High Energy Physics, Tampere, Finland, 
July 15--21, 1999.

\bibitem{KR} J.~K\"uhn and G.~Rodrigo, \Journal{\PRL}{81}{49}{1998};
\Journal{\PRD}{59}{054017}{1999}.

\bibitem{TZC} CDF Collaboration, F.~Abe {\it et al.},
\Journal{\PRL}{80}{2525}{1998}.

\bibitem{EHS} G.~Eilam, J.~Hewett, and A.~Soni, 
\Journal{\PRD}{44}{1473}{1991}; Erratum, {\it ibid.} {\bf 59}, 039901 (1999). 

\bibitem{ATLAS} ATLAS Technical Design Report, Vol.~II, CERN/LHCC/99-15 (1999).

\bibitem{YELLOW} M.~Beneke {\it et al.}, hep-ph/0003033.

\bibitem{MuP} H.~Murayama and M.~Peskin, 
{\em Ann.~Rev.~Nucl.~Part.~Sci.}~{\bf 46}, 533 (1996).

\bibitem{SMW} A.~Stange, W.~Marciano, and S.~Willenbrock, \Journal{\PRD}{49}
{1354}{1994}; {\bf 50}, 4491 (1994).

\bibitem{HZ} T.~Han and R.-J.~Zhang, \Journal{\PRL}{82}{25}{1999}.

\bibitem{HTZ} T.~Han, A.~Turcot, and R.-J.~Zhang, 
\Journal{\PRD}{59}{093001}{1999}.

\bibitem{BHT} H.~Baer, B.~Harris, and X.~Tata, \Journal{\PRD}{59}{015003}
{1999}.

\bibitem{CMW} M.~Carena, S.~Mrenna, and C.~Wagner, 
\Journal{\PRD}{60}{075010}{1999}.

\end{thebibliography}
